\documentclass[aps,prd,twocolumn,superscriptaddress,nofootinbib,10pt]{revtex4-1}
\usepackage[paperwidth=21cm,paperheight=29.7cm,top=2.54cm,bottom=2.54cm,left=2cm,right=2cm]{geometry}
\usepackage[colorlinks,linkcolor=blue,anchorcolor=blue,citecolor=blue,urlcolor=blue]{hyperref}
\usepackage{graphicx,subfigure}
\usepackage[figuresright]{rotating}
\usepackage{amsmath,amsfonts,amssymb,bm}
\usepackage{array,enumitem,multirow}
\usepackage{acronym}
\usepackage{makecell}
\newcommand{\be}{\begin{equation}}
\newcommand{\ee}{\end{equation}}
\newcommand{\bea}{\begin{eqnarray}}
\newcommand{\eea}{\end{eqnarray}}

\newacro{GR}{general relativity}
\newacro{GW}{gravitational wave}
\newacro{MG}{modified gravity theory}
\newacro{BH}{Black hole}
\newacro{PN}{post-Newtonion}
\newacro{ppE}{parameterized post-Einsteinian}
\newacro{GCB}{galactic ultra-compact binary}
\newacro{SBHB}{stellar-mass black hole binary}
\newacro{MBHB}{massive black hole binary}
\newacro{BHB}{black hole binary}
\newacro{IMBHB}{intermediate-mass black hole binary}
\newacro{EMRI}{extreme mass ratio inspiral}
\newacro{IMRI}{intermediate mass ratio inspiral}
\newacro{SGWB}{stochastic gravitational wave background}
\newacro{MECO}{minimal energy circular orbit}
\newacro{FAR}{false alarm rate}
\newacro{CE}{Cosmic Explorer}
\newacro{ET}{Einstein Telescope}
\newacro{LISA}{Laser Interferometer Space Antenna}
\newacro{EdGB}{Einstein-dilaton Gauss-Bonnet}
\newacro{dCS}{dynamic Chern-Simons}
\newacro{SNR}{signal-to-noise ratio}
\newacro{FIM}{Fisher Information Matrix}
\newacro{ISCO}{innermost stable circular orbit}
\newacro{NSBH}{neutron star-black hole binary}
\newacro{MCMC}{Markov Chain Monte Carlo}
\newacro{QNM}{quasi-nomral mode}
\newcommand{\ma}{\mathcal{A}}

\begin{document}

\title{ Quasinormal modes of accelerating spacetime}

\author{Tao Zhou}
 \email{202220130252@mail.scut.edu.cn}%
\author{Peng-Cheng Li}%
 \email{pchli2021@scut.edu.cn, corresponding author}
\affiliation{%
School of Physics and Optoelectronics, South China University of Technology, Guangzhou 510641, People’s Republic of China.
}%

\date{\today}

\begin{abstract}
We calculate the exact values of the quasinormal frequencies for massless perturbations with spin $s\leq2$ moving in pure accelerating spacetime. We use two different methods to transfer the perturbation equations into the form of hypergeometric differential equations and obtain the same quasinormal frequencies. These purely imaginary spectra are shown to be independent of the spin of the perturbation and match those of the  so-called  acceleration modes of accelerating black holes after taking the Minkowski limit. This implies that the acceleration modes  actually originate from the pure accelerating spacetime and the appearance of black holes would deform the spectra. In addition, we calculate the quasinormal frequencies of scalar, electromagnetic and gravitational perturbations of $D$-dimensional de Sitter spacetime and compare them with previous results to verify the validity of our method.
\end{abstract}

\maketitle


\section{\label{sec:level1}INTRODUCTION}
The first detection of gravitational waves (GWs) from the merger of two black holes (BHs) \cite{LIGOScientific:2016aoc} and the release of the first image of the supermassive BH in M87 \cite{EventHorizonTelescope:2019dse} have further convinced people that BHs are real celestial objects in the universe, not just theoretical concepts. When a BH gets perturbed, the relaxation can be described by a superposition
of exponentially damped sinusoids termed quasinormal modes (QNMs) \cite{QNMKokkotas_1999,QNMKonoplya_2011,QNMBERTI}. Thus, in the ringdown stage of the coalescence of two astrophysical BHs, the GWs take the form of superposed QNMs of
the remnant BH. Due to the no-hair theorem, the frequencies and decay rates of these QNMs are uniquely determined by the final BH' s physical parameters \cite{Bekenstein:1996pn}. The measurement of the QNMs from GW observations would help us to test general relativity and probe the nature of remnants
from compact binary mergers. This program is called BH spectroscopy \cite{Dreyer:2003bv}. Moreover, the QNMs can also be used to determine the linear stability of a perturbed BH. For example, the study of QNMs of massless scalar fields in the exterior of Reissner-Nordstr\"om-de Sitter (RNdS) BHs can be used to determine whether the strong cosmic censorship conjecture is violated \cite{Cardoso:2017soq}. 
 
Recently, the QNMs of the so-called C-metric in various situations have garnered some attention \cite{Nozawa:2008wf, Destounis:2020pjk,Fontana:2022whx, Destounis:2022rpk,Gwak:2022nsi,Xiong:2023usm,Lei:2023mqx,BarraganAmado:2023wxt,Chen:2024rov}. The C-metric describes an axially symmetric and  stationary spacetime that contains two causally separated  BHs accelerating away from each other in opposite spatial directions  under the action of conical singularities  along the axis \cite{Kinnersley:1970zw,Plebanski:1976gy,Griffiths:2006tk}. This model might be useful for understanding the behavior of moving and accelerating BHs, such as those resulting from a BH superkick or a cosmic string connecting two BHs \cite{Hawking:1995zn,Merritt:2004xa}. These studies of QNMs of accelerating BHs have taken advantage of the C-metric being of Petrov type D, which allows the perturbation equations for various test fields to be separable \cite{Hawking:1997ia,Bini:2008mzd,Kofron:2015gli,Frolov:2018pys}. One interesting feature is the QNMs of both the charged accelerating BHs and the rotating accelerating BHs can be classified into three families: the photon sphere modes, the near-extreme modes and the acceleration modes. The photon sphere modes are associated with the peaks of the potential barrier, while the near
extreme modes are related to the near-horizon geometry of the BHs. The acceleration modes depend solely on the acceleration horizon  and are absent in non-accelerating spacetimes.  This new kind of modes was first identified for scalar perturbations  of the charged accelerating BHs \cite{Destounis:2020pjk}.  Later, the similar phenomenon was also verified for scalar \cite{Xiong:2023usm} and gravitational perturbations \cite{Chen:2024rov} of the rotating accelerating BHs. 

In this study, we aim to explore the origin of the acceleration modes. Given that the acceleration modes have a weak dependence on the BH's charge or rotational parameter and depend solely on the surface gravity of the Minkowski limit rather than the BH' s surface gravity, we naturally speculate that the acceleration modes might originate from a pure accelerating spacetime without a BH. It is worth noting that in pure de Sitter (dS) spacetime, there exist dS  modes \cite{Lopez-Ortega:2006aal,Lopez-Ortega:2006tjo,Lopez-Ortega:2012xvr}, which become deformed when BHs appear \cite{Jansen:2017oag, Cardoso:2017soq,Destounis:2018qnb, Konoplya:2022xid}. Due to the remarkable similarity between the accelerating spacetime and the dS spacetime that the former has a accelerating horizon and the latter has a cosmological horizon, it is reasonable to connect the acceleration modes with the counterpart in the pure accelerating spacetime.

We would like to calculate the QNMs of massless perturbations with spin $s\leq2$ in pure accelerating spacetime. By taking the Minkowski limit of the master equations describing perturbations of accelerating BHs---specifically by setting the mass, electric charge, or spin to zero---we can derive the equations that govern perturbations in the empty accelerating spacetime. Two methods are employed to solve the perturbation equations. The first one is to follow the approach in \cite{Lopez-Ortega:2006aal}, via direct coordinate transformation  we can  find that the perturbation equations are of hypergeometric type. Then the QNMs are easily obtained when the proper boundary conditions are imposed. This method works straightforward for scalar perturbations \cite{Destounis:2020pjk}. The other approach is to analyze the properties of the Minkowski limit of the Teukolsky-like equations that govern various massless field perturbations of spinning accelerating BHs \cite{Bini:2008mzd}. This involves identifying all singular points and demonstrating that the equations are of the Fuchsian type, with all singular points being regular singular points. If exactly three regular singular points are present, the equations can be simply transformed into the standard form of the hypergeometric differential equations (HDEs)  \cite{wang1989special,gray2008linear}.  Then imposing the proper boundary conditions we can derive the QNMs. To validate our approach, we employ this method to calculate dS modes and compare our results with those in \cite{Lopez-Ortega:2006aal}. We confirm that our results are identical for dS modes of scalar, electromagnetic, and gravitational perturbations. This advantage of this method is that it does not require  complex coordinate transformations to convert the  equations into the form of HDEs.

The remainder of this paper is organized as follows. In Sec. \ref{sec:level2}, we explicitly show the derivation of QNMs of massless scalar perturbations in pure accelerating spacetime by solving the Minkowski limit of the Klein-Gordon (KG) equation in the background of the C-metric, employing the two methods described above. We calculate the QNMs of massless perturbations with spin $s\leq2$ in Sec. \ref{sec:level3} by solving the Minkowski limit of the Teukolsky-like equations in the spinning C-metric spacetime. Moreover, in Sec. \ref{sec:level4} we use the new method to calculate dS modes of scalar, electromagnetic, and gravitational perturbations of $D$ dimensional dS spacetime and compare the results with those in   \cite{Lopez-Ortega:2006aal}.  We give a brief summary in Sec. \ref{sec:level5}. Finally, in Appendix \ref{appdixFeq} we introduce some basics of Fuchsian equations involved in this work. By convention, we employ geometric units $c=G=1$ and the metric signature $(-,+,+,+)$.
\section{\label{sec:level2}Scalar Field}
We first study the QNMs of scalar perturbations of pure accelerating spacetime.  We consider a massless neutral scalar filed minimally coupled to gravity living in the spacetime of a C-metric. The evolution of the scalar perturbations is described by the KG equation, which has been shown to be separable through conformal transformation, as demonstrated in \cite{Destounis:2020pjk}. The C-metric  that covers one BH  can be written in terms of spherical-type coordinates as \cite{Griffiths:2006tk}
\begin{equation}\label{chargedCmetric}
	\begin{aligned}
		d s^2= & \frac{1}{(1-\ma r \cos \theta)^2}\left(-f(r) d t^2+\frac{d r^2}{f(r)}\right. \\
		 & \left.+\frac{r^2 d \theta^2}{P(\theta)}+P(\theta) r^2 \sin ^2 \theta d \varphi^2\right),
	\end{aligned}
\end{equation}
where
\begin{eqnarray}
		 f(r)&=&\left(1-\frac{2 M}{r}\right)\left(1-\ma^2 r^2\right),\\
        P(\theta)&=&1-2 \ma M \cos \theta.
\end{eqnarray}
Here $\ma$ is the parameter describing acceleration, $M$ is the mass of the BH. As  $\ma\to0$, this metric asymptotes to  Schwarzschild metric. As explained in \cite{Griffiths:2009dfa}, taking the Minkowski limit $M=0$, the metric (\ref{chargedCmetric}) in the region $r<1/\ma$ can be turned into the uniformly accelerated metric (also named {\em Rindler} metric) through an appropriate coordinate transformation and $\ma$ is interpreted as the acceleration of a test particle located at the origin $r=0$.

Ref. \cite{Destounis:2020pjk} found that under the conformal transformation $\widetilde{g}_{\mu\nu}\to\Omega^2g_{\mu\nu}$, $\widetilde{\Phi}\to\Omega^2\Phi$, where $\Omega=1-\ma r \cos \theta$, the KG equation $\nabla^\mu\nabla_\mu\Phi=0$ becomes separable 
by choosing the ansatz 
\begin{equation}\label{conformal}
	\widetilde{\Phi}=\sum _{lm}e^{-i \omega_{lm} t} e^{i m \varphi} \frac{\phi_{lm}(r)}{r} \chi_{lm}(\theta),
\end{equation}
where $\omega_{lm}$ is the quasinormal frequency and $m$ is the magnetic (azimuthal) quantum number. In the following we will neglect the subscript $lm$ for simplicity. As one can see from the metric (\ref{chargedCmetric}) that the conical singularities occur  on the axis at both $\theta=0$ and $\theta=\pi$.  To remove the 
conical singularity at $\theta=\pi$, one can specify the period of the azimuthal coordinate $\varphi$ as $2\pi C$, where $C=1/P(\pi)$. As a consequence, the magnetic quantum number $m$ would not be an integer, instead, $m$ must be of the form $m=m_0 P(\pi)$ with $m_0$ being an  integer \cite{Bini:2008mzd}, since one has $e^{im(\varphi+2\pi C)}=e^{im\varphi}$. 

Consequently, one has the following separated equations
	\begin{eqnarray}\label{eq:scmain1}
		\frac{d^2 \phi(r)}{d r_*^2}+\left(\omega^2-V_r\right) \phi(r)&=&0,\\
		\frac{d^2 \chi(\theta)}{d z^2}-\left(m^2-V_\theta\right) \chi(\theta)&=&0,\label{eq:scmain2}
	\end{eqnarray}
where
\begin{equation}
	d r_*=\frac{d r}{f(r)}, \quad d z=\frac{d \theta}{P(\theta) \sin \theta},
\end{equation}
and
	\begin{eqnarray}
		V_r&=&f(r)  \left(\frac{\lambda}{r^2}-\frac{f(r)}{3 r^2}+\frac{f^{\prime}(r)}{3 r}-\frac{f^{\prime \prime}(r)}{6}\right), \\
			V_\theta&=&P(\theta)  \left(\lambda \sin ^2 \theta-\frac{P(\theta) \sin ^2 \theta}{3}+\frac{\sin \theta \cos \theta P'(\theta)}{2}\right.\nonumber \\
			& &\left.+\frac{\sin ^2 \theta P''(\theta)}{6}\right).
	\end{eqnarray}
Here $r_*$ is the tortoise coordinate and $\lambda$ is a separation constant. Taking the Minkowski limit $M\rightarrow0$, the separated wave equations \eqref{eq:scmain1} and \eqref{eq:scmain2} that govern the scalar perturbations in pure accelerating spacetime are reduced to
\begin{eqnarray}\label{eq:scrad}
	&&\left(1-\ma^2 r^2\right) \frac{d}{d r}\left[\left(1-\ma^2 r^2\right) \frac{d \phi(r)}{d r}\right]\nonumber\\
	&&\quad +\left[\omega^2-\left(1-\ma^2 r^2\right)\left(\frac{\lambda}{r^2}-\frac{1}{3r^2}\right)\right] \phi(r)=0,
	\end{eqnarray}
and
\begin{equation}\label{eq:scmag}
	\sin \theta \frac{d}{d \theta}\left(\sin \theta \frac{d \chi(\theta)}{d \theta}\right)+\left[(\lambda-\frac{1}{3} )\sin ^2 \theta-m^2\right] \chi(\theta)=0.
\end{equation}
Note that in the Minkowski limit there is no difference for $m$ and $m_0$. 
We can see that the angular equation (\ref{eq:scmag}) has the same form as the equation of the Laplacian spherical harmonics $Y_{lm}$ in spherical coordinates. Therefore, we can write the angular equation as 
\begin{equation}
	\sin \theta \frac{d}{d \theta}\left(\sin \theta \frac{d \chi(\theta)}{d \theta}\right)+\left[\ell(\ell+1) \sin ^2 \theta-m^2\right] \chi(\theta)=0,
\end{equation}
from which we find the separation constant $\lambda$ is related to the eigenvalue of the spherical Laplacian operator by
\begin{equation}
	\lambda=\ell(\ell+1)+\dfrac{1}{3}.
\end{equation}
\subsection{Direct conversion to HDEs}
We follow the steps in \cite{Lopez-Ortega:2006aal} to show that the radial equation \eqref{eq:scrad} can be transformed into  the HDE by a change of variable. First of all, by introducing two dimensionless quantities 
\begin{equation}
x=r \ma,\quad {\rm and}\quad \omega=\ma\,\widetilde{\omega},
\end{equation}
 the equation \eqref{eq:scrad} can be written as
\begin{equation}\label{scalarxeq}
	\left(x^2-1\right) \frac{d^2 \phi}{d x^2}+2 x \frac{d \phi}{d x}+\left(\frac{\widetilde{\omega}^2}{x^2-1}+\frac{\ell(\ell+1)}{x^2}\right) \phi=0,
\end{equation}
Then making the change of variable\ $y=x^2$, the equation is transformed into 
\begin{equation}
	4 y(y-1) \frac{d^2 \phi}{d y^2}+(6 y-2) \frac{d \phi}{d y}+\left(\frac{\ell(\ell+1)}{y}+\frac{\widetilde{\omega}^2}{y-1}\right) \phi=0.
\end{equation}
Furthermore, we make the ansatz 
\begin{equation}\label{ansatzphi}
\phi=y^A (1-y)^B \tilde{\phi},
\end{equation}
where the parameter $A$ and $B$ are chosen the following values
\begin{equation}
	A=\left\{\begin{array}{l}
		\frac{\ell+1}{2}, \\ \\
		-\frac{\ell}{2},
	\end{array} \qquad B= \pm \frac{i \widetilde{\omega}}{2},\right.
\end{equation}
to find that the function $\tilde{\phi}$ must be the solution of the HDE
\begin{equation}\label{hyper1}
	y(1-y)\frac{d^2 \tilde{\phi}}{d y^2}+\left[\gamma-(\alpha+\beta+1) y\right] \frac{d \tilde{\phi}}{d y} -\alpha \beta \tilde{\phi}=0
\end{equation}
with the parameters $\alpha$, $\beta$ and $\gamma$ are equal to 
\begin{equation}\label{abcvalues}
	\begin{aligned}
		\alpha & =A+B, \\
		\beta & =A+B+\frac{1}{2}, \\
		\gamma & =2 A+\frac{1}{2} .
	\end{aligned}
\end{equation}
One can verify that Eq. \eqref{hyper1} indeed has three regular singular points $y=0$, $1$ and $\infty$, which is of the standard form of the HDE. In contrast, Eq. \eqref{scalarxeq} has three regular singular points $x=0$, $1$ and $-1$, while $\infty$ is an ordinary point. The structure of the regular singular points is altered by the change of variable from $x$ to $y$. The further transformation \eqref{ansatzphi} is employed to obtain the standard form of the HDE. There are several possible combinations for the values of $A$ and $B$ (and therefore of $\alpha$, $\beta$, and $\gamma$). Thus the solutions of Eq. \eqref{hyper} can be written in
several equivalent forms. In the following we study in detail  the case $A=\frac{\ell+1}{2}$ and $B=\frac{i \widetilde{\omega}}{2}$.

Similar to dS QNMs, the QNMs are solutions to the equations of motion that satisfy the physical boundary conditions: (i) the field is regular at the origin; (ii) the field is purely outgoing near the acceleration horizon. Depending on the values of the parameters, the HDEs have different solutions \cite{Lopez-Ortega:2006aal,abramowitz1965handbook}. 
The regular behavior at $r=0$ singles out the solution of  Eq. \eqref{scalarxeq} to be
\begin{equation}\label{phisolution}
		{\phi}=y^{\frac{\ell+1}{2}}(1-y)^{\frac{i \widetilde{\omega}}{2}}{ }_2 F_1(\alpha, \beta; \gamma; y),
\end{equation}
where ${ }_2 F_1(\alpha, \beta; \gamma; y)$ denotes the hypergeometric
function. To explore the behavior of the solution at $y\to1$, namely the acceleration horizon, we utilize the linear transformation formulas for the hypergeometric
functions \cite{abramowitz1965handbook}. 
If the quantity $\gamma-\alpha-\beta=-i \widetilde{\omega}$ is not an integer, then the relation between two hypergeometric
functions with variables $y$ and $1-y$ is 
\begin{equation}
	\begin{aligned}
		{\phi} & =\frac{\Gamma(\gamma) \Gamma(\gamma-\alpha-\beta)}{\Gamma(\gamma-\alpha) \Gamma(\gamma-\beta)}y^{\frac{\ell+1}{2}}(1-y)^{\frac{i \widetilde{\omega}}{2}} \\
		&\quad \times{ }_2 F_1(\alpha, \beta; \alpha+\beta-\gamma+1 ; 1-y) \\
		&\quad +\frac{\Gamma(\gamma) \Gamma(\alpha+\beta-\gamma)}{\Gamma(\alpha) \Gamma(\beta)}y^{\frac{\ell+1}{2}}(1-y)^{-\frac{i \widetilde{\omega}}{2}} \\
		&\quad \times{ }_2 F_1(\gamma-\alpha, \gamma-\beta ; \gamma-\alpha-\beta+1 ; 1-y),
	\end{aligned}
\end{equation}
where $\Gamma(y)$ denotes the Gamma function. We can check that the first term on the right hand side of above equation represents an ingoing wave and the second term represents an outgoing wave. To satisfy the boundary conditions of QNMs, we should discard the first term. This is achieved when we impose \footnote{Note that repeated symbols in each subsection of this paper may represent different quantities.}
\begin{equation}\label{bd0}
	\gamma-\alpha=-n, \quad \text { or } \quad \gamma-\beta=-n, \quad n=0,1,2, \cdots,
\end{equation}
such that $\Gamma(\gamma-\alpha)$ or $\Gamma(\gamma-\beta)$ becomes infinity.
However, from \eqref{abcvalues} we can find that if  $ \gamma-\alpha-\beta=-i \widetilde{\omega}$ is not an integer, then neither $\gamma-\alpha=(\ell+2-i\widetilde{\omega})/2$ nor $\gamma-\beta=(\ell+1-i\widetilde{\omega})/2$ can be an integer, since $(\ell+2)/2$ and $(\ell+1)/2$ are integers or half-integers, $-i\widetilde{\omega}/2$ is not an integer or half-integer and
therefore their sum cannot be an integer. Hence the condition \eqref{bd0} cannot be satisfied if $ \gamma-\alpha-b$ is not an integer. 

Instead, we must assume  $\gamma-\alpha-\beta$  is an integer. If $\gamma-\alpha-\beta=-n_1$, where\ $ n_1=1,2,3,...$ \footnote{We do not need to consider $\gamma-\alpha-\beta$ is equal to zero or a positive value, since the former means the scalar perturbations are always static and the latter makes the scalar perturbations unstable. }, we can write the radial function \eqref{phisolution} as
\begin{equation}
	\begin{aligned}
		& \phi=y^{\frac{\ell+1}{2}}(1-y)^{\frac{i \widetilde{\omega}}{2}}\left\{\frac{\Gamma\left(\alpha+\beta-n_1\right) \Gamma\left(n_1\right)}{\Gamma(\alpha) \Gamma(\beta)}(1-y)^{-n_1}\right. \\
		& \times \sum_{s=0}^{n_1-1} \frac{\left(\alpha-n_1\right)_s\left(\beta-n_1\right)_s}{s !\left(1-n_1\right)_s}(1-y)^s\\
		&-\frac{(-1)^{n_1} \Gamma\left(\alpha+\beta-n_1\right)}{\Gamma\left(\alpha-n_1\right) \Gamma\left(\beta-n_1\right)} \\
		& \times \sum_{s=0}^{\infty} \frac{(\alpha)_s(\beta)_s}{s !(n+s) !}(1-y)^s[\ln (1-y)-\psi(s+1) \\
		&- \psi(s+n+1)+\psi(\alpha+s)+\psi(\beta+s)]\},
	\end{aligned}
\end{equation}
where $\psi(y)=d\Gamma(y)/dy$. We can see that the first term  in curly brackets represents an outgoing wave  while the second term represents an ingoing wave. Thus to satisfy the boundary
conditions of QNMs, we must impose the condition
\begin{equation}\label{bound}
	\alpha-n_1=-n,  \text { or } \,\, \beta-n_1=-n, \quad n=0,1,2, \cdots,
\end{equation}
to retain the outgoing wave. Combine this with $\gamma-\alpha-\beta=-n_1$, we can find the accelerating QN frequencies are equal to
\begin{equation}
	i \widetilde{\omega}=\ell+1+2 n, \quad i \widetilde{\omega}=\ell+2+2 n,
\end{equation}
which can also be written as 
\begin{equation}\label{iomegascalar1}
i \widetilde{\omega}= \ell+\widetilde{n},\quad \widetilde{n}=1,2,3,\cdots.
\end{equation}
The result shows that the previous assumption is self-consistent. We can verify that our result matches those of the acceleration modes for a scalar field moving in the spacetime of the charged C-metric \cite{Destounis:2020pjk} and spinning C-metric \cite{Xiong:2023usm} when taking the Minkowski limit. Therefore, we may arrive at the conclusion that the acceleration modes found for accelerating BHs in fact originate from the empty accelerating spacetime and get deformed when BHs present in the spacetime.

\subsection{Fuchsian equations}
In this subsection, we will derive the QNMs of the scalar field in the empty accelerating spacetime by using the simple connection between HDEs and Fuchsian equations. A linear differential equation in which every singular point, including the point at infinity, is a regular singularity is called Fuchsian equation or equation of Fuchsian type. In particular, a Fuchsian equation with three regular singular points reduces to the HDE. In the appendix \ref{appdixFeq} we give a brief review of the basics  of Fuchsian equations relevant to this work by referring to Refs. \cite{wang1989special} and \cite{gray2008linear}. The knowledge of Fuchsian equations can help us  transform an equation with three arbitrary regular singular points into the form of the standard HDE in a programmatic way. 

First of all, the radial equation \eqref{scalarxeq} can be cast  into the form of Eq. \eqref{eqfx} with
\begin{eqnarray}
	p(x)&=&\frac{2x}{x^2-1}, \\
	q(x)&=&\frac{\widetilde{\omega}^2 x^2+\ell(\ell+1)(x^2-1)}{(x^2-1)^2x^2}.
\end{eqnarray}
We can easily identify the three  regular singular points, which are denoted by $a_1=0$, $a_2=1$ and $a_3=-1$. Then from Eqs. \eqref{eq:px} and \eqref{eq:qx} by expanding $p(x)$ and $q(x)$ around these singular points, we can determine the coefficients 
\begin{eqnarray}
	A_1&=&0,\quad B_1=-\ell(\ell+1),\quad C_1=0,\\
	A_2&=&1,\quad B_2=\frac{\widetilde{\omega}^2}{4},\quad C_2=\frac{-\widetilde{\omega}^2+2\ell(\ell+1)}{4},\\
	A_3&=&1,\quad B_3=\frac{\widetilde{\omega}^2}{4},\quad C_3=\frac{\widetilde{\omega}^2-2\ell(\ell+1)}{4}.
\end{eqnarray}
Since $x=\infty$ is an ordinary point, the parameters $A_r$, $B_r$ and $C_r$, with $r=1, 2, 3$, should satisfy the constraint \eqref{xinftyeq}. We can check from above expressions that it is indeed this case. 

Next, from the indicial equation \eqref{eq:index}, we can determine the characteristic exponents of each regular singular point, which are given by
\begin{eqnarray}
&&\alpha_1=\ell+1,\quad \alpha_2=-\ell, \nonumber\\ &&\beta_1=\frac{i\widetilde{\omega}}{2},\quad \beta_2=-\frac{i\widetilde{\omega}}{2},\\
&&\gamma_1=\frac{i\widetilde{\omega}}{2},\quad
\gamma_2=-\frac{i\widetilde{\omega}}{2}.\nonumber
\end{eqnarray}
One can check that above result satisfies the constraint \eqref{alphasbeta1}. Through the transformation \eqref{apa1}, that is
\begin{equation}
	\begin{aligned}
		& y=\frac{2x}{x+1},\\
	&\phi=\left(\frac{x}{x+1}\right)^{\ell+1}\left(\frac{x-1}{x+1}\right)^{\frac{i\widetilde{\omega}}{2}} g(y),
\end{aligned}
\end{equation}
 the radial equation \eqref{scalarxeq} can be turned into the standard form of HDE \eqref{hyper}, with the parameters
given by \eqref{indexrel} 
\begin{equation}
	\begin{aligned}
		& \alpha=\ell+1+i\widetilde{\omega}, \\
		& \beta=\ell+1, \\
		& \gamma=2\ell+2.
	\end{aligned}
\end{equation}
As discussed in the previous subsection, the one of the solutions of Eq. \eqref{scalarxeq} that is regular at $r=0$ is given by
\begin{equation}\label{radialf2}
	{\phi}=\left(\frac{y}{2}\right)^{\ell+1}(y-1)^{\frac{i \widetilde{\omega}}{2}}{ }_2 F_1(\alpha, \beta ; \gamma ; y).
\end{equation}
As before, we need to change the variable of hypergeometric function from $y$ to $1-y$ to explore the behavior of the solution at the acceleration horizon. 
If the quantity $\gamma-\alpha-\beta=-i \widetilde{\omega}$ is not an integer, then we have
\begin{equation}
	\begin{aligned}\label{noninterger}
		{\phi} & =\frac{\Gamma(\gamma) \Gamma(\gamma-\alpha-\beta)}{\Gamma(\gamma-a) \Gamma(\gamma-\beta)}\left(\frac{y}{2}\right)^{\ell+1}(y-1)^{\frac{i \widetilde{\omega}}{2}} \\
		&\quad \times{ }_2 F_1(\alpha, \beta; \alpha+\beta-\gamma+1 ; 1-y) \\
		&\quad +\frac{\Gamma(\gamma) \Gamma(\alpha+\beta-c)}{\Gamma(\alpha) \Gamma(\beta)}\left(\frac{y}{2}\right)^{\ell+1}(1-y)^{-\frac{i \widetilde{\omega}}{2}} (-1)^{\frac{i \widetilde{\omega}}{2}} \\
		&\quad \times{ }_2 F_1(\gamma-\alpha, \gamma-\beta; \gamma-\alpha-\beta+1 ; 1-y).
	\end{aligned}
\end{equation}
The second term on the right hand side of above equation represents an outgoing wave, which is allowed by the boundary condition at the acceleration horizon. To achieve this, we should impose
\begin{equation}
	\gamma-\alpha=-n, \quad \text { or } \quad \gamma-\beta=-n, \quad n=0,1,2, \cdots.
\end{equation}
One can check that this condition cannot be satisfied, which means we should instead assume that  $\gamma-\alpha-\beta$ is an integer.  If $\gamma-\alpha-\beta=-n_1$, where\ $ n_1=1,2,3,...$ , the solution can be written as
\begin{equation}\label{integer}
	\begin{aligned}
		& \phi=\left(\frac{y}{2}\right)^{\ell+1}(y-1)^{\frac{i \widetilde{\omega}}{2}}\left\{\frac{\Gamma\left(\alpha+\beta-n_1\right) \Gamma\left(n_1\right)}{\Gamma(\alpha) \Gamma(\beta)}(1-y)^{-n_1}\right. \\
		& \times \sum_{s=0}^{n_1-1} \frac{\left(\alpha-n_1\right)_s\left(\beta-n_1\right)_s}{s !\left(1-n_1\right)_s}(1-y)^s-\frac{(-1)^{n_1} \Gamma\left(\alpha+\beta-n_1\right)}{\Gamma\left(\alpha-n_1\right) \Gamma\left(\beta-n_1\right)} \\
		& \times \sum_{s=0}^{\infty} \frac{(\alpha)_s(\beta)_s}{s !(n+s) !}(1-y)^s[\ln (1-y)-\psi(s+1) \\
		&- \psi(s+n+1)+\psi(\alpha+s)+\psi(\beta+s)]\}.
	\end{aligned}
\end{equation}
We can see that the first term  in curly brackets represents an outgoing wave  while the second term represents an ingoing wave. Thus to satisfy the boundary conditions of QNMs, we must impose 
\begin{equation}\label{bound}
	\alpha-n_1=-n,  \text { or } \,\, \beta-n_1=-n, \quad n=0,1,2, \cdots.
\end{equation}
Combine this with $\gamma-\alpha-\beta=-n_1$, the accelerating QN frequencies are equal to 
\begin{equation}
	i \widetilde{\omega}= \ell+n+1,
\end{equation}
which is equivalent to \eqref{iomegascalar1}. In this way we get the identical result using two methods, although the radial functions \eqref{phisolution} and \eqref{radialf2} in these two cases are different.
\section{Massless perturbations of any spin}\label{sec:level3}
In this section we will calculate the QNMs of massless perturbations of any spin of the accelerating spacetime. 
The master equation describing  massless perturbations of spinning C-metric due to fields of any spin has been derived in \cite{Bini:2008mzd} by following the approach of Teukolsky \cite{Teukolsky:1973ha} within the context of the Newman-Penrose formalism \cite{Newman:1961qr}. Remarkably, similar to the Teukolsky equation of the Kerr BH, the master equation  can be separated into its radial and angular parts, due to the fact that the spinning C-metric is of Petrov type D. By taking the Minkowski limit, we can obtain the equations capturing the dynamics of scalar, Dirac,   electromagnetic  and gravitational perturbations in the empty accelerating spacetime.

The metric of spinning C-metric in terms of  the Boyer-Lindquist-type coordinates $(t, r, \theta, \phi)$ is given by \cite{Griffiths:2005se}
\begin{eqnarray}
		ds^2&= & \frac{1}{\Omega^2}\Bigg\{-\frac{1}{\Sigma}\left(Q-a^2 P \sin ^2 \theta\right) d t^2\nonumber\\
			&&+\frac{2 a \sin ^2 \theta}{\Sigma}\left[Q-P\left(r^2+a^2\right)\right] d t d \phi\nonumber \\
	&& +\frac{\sin ^2 \theta}{\Sigma}\left[P\left(r^2+a^2\right)^2-a^2 Q \sin ^2 \theta\right] d \phi^2\nonumber\\
	&&+\frac{\Sigma}{Q} d r^2+\frac{\Sigma}{P} d \theta^2\Bigg\},
\end{eqnarray}
where $a$ is the rotation parameter and the functions $\Omega$, $\Sigma$, $P$ and $Q$ are defined by
\begin{equation}
	\begin{aligned}
		& \Omega=1-\ma r \cos \theta, \quad \Sigma=r^2+a^2 \cos ^2 \theta, \\
		& P=1-2 \ma M \cos \theta+a^2 \ma^2 \cos ^2 \theta, \\
		& Q=\Delta\left(1-\ma^2 r^2\right), \quad \Delta=r^2-2 M r+a^2.
	\end{aligned}
\end{equation}
The master equation describing the dynamics of a massless field with spin weight $s$ in this spacetime is given by 
\begin{equation}
	\begin{gathered}\label{masteeq}
		{\left[\left(\nabla^\mu-s \Gamma^\mu\right)\left(\nabla_\mu-s \Gamma_\mu\right)+4 s^2 \Psi_2\right] \psi=0}, \\
		s=0, \pm \frac{1}{2}, \pm 1, \pm \frac{3}{2}, \pm 2,
	\end{gathered}
\end{equation}
where $\Psi_2=-(1+ia\ma)M\Omega^3/(r-ia\cos \theta)^3$ is the nonvanishing  Weyl scalar in the spinning C-metric background and  a ``connection vector'' is introduced by
\begin{eqnarray}
	\Gamma^t&=&\frac{\Omega^2}{\Sigma}\Bigg\{\frac{1}{Q^2}[M(\ma^2r^4+a^2)+r(1+a^2\ma^2)(\Delta-Mr)]\nonumber\\
	&&+i\frac{a}{P}[(1+a^2\ma^2)\cos\theta-\ma M(1+\cos^2\theta)] \Bigg\},\nonumber\\
	\Gamma^r&=&-\frac{\Omega}{\Sigma}\left(\frac{1}{2}\Omega\partial_rQ+2\ma \cos\theta Q\right),\nonumber\\
	\Gamma^\theta&=&\frac{2\ma \Omega P r \sin \theta}{\Sigma},\nonumber\\
		\Gamma^\phi &=& -\frac{\Omega^2}{\Sigma}\Bigg[ \frac{a \partial_rQ}{2Q}+i\frac{\cos\theta(2P-1)}{P\sin^2\theta}\nonumber\\
		&&\qquad+i\frac{\ma M(\cos^2\theta-\ma^2 a^2 \cos\theta+1)}{P\sin^2\theta}\Bigg].
\end{eqnarray}
Ref. \cite{Bini:2008mzd} demonstrated that the master equation \eqref{masteeq} admits separable solutions of the form
\begin{equation}
	\psi(t, r, \theta, \phi)=\sum_{lm}\Omega^{(1+2 s)} e^{-i \omega_{lm} t} e^{i m \phi} R_{lm}(r) S_{lm}(\theta),
\end{equation}
where $\omega_{lm}$ is the wave frequency and $m$ is the azimuthal number. Again, we will neglect  the subscript $lm$ for simplicity in the following discussions.
The radial equation is then
\begin{equation}
	Q^{-s}\frac{d}{dr}\left(Q^{s+1}\frac{dR(r)}{dr}\right)+V_{\rm (rad)}R(r)=0,
\end{equation}
with
\begin{eqnarray}
	V_{\rm (rad)}&=&-2r\ma^2(r-M)(1+s)(1+2s)\nonumber\\
	&&+\frac{((r^2+a^2)\omega-a m)^2}{Q}-2is\Bigg[-\frac{a m \partial_rQ}{2Q}\nonumber\\
	&&+\frac{\omega M(r^2-a^2)}{\Delta}-\frac{\omega r\sigma_0}{1-\ma^2r^2}\Bigg]+2K,
\end{eqnarray}
where $K$ is the separation constant and $\sigma_0=(1+a^2\ma^2)$. By introducing
\begin{equation}
	H(r)=(r^2+a^2)^{1/2}Q^{s/2},
\end{equation}
and the ``tortoise'' coordinate $r_*$, where
\begin{equation}
	\frac{dr}{dr_*}=\frac{Q}{r^2+a^2},
\end{equation}
 the radial equation can be transformed into the one-dimensional Schr\"odinger-like equation
\begin{equation}\label{rdeq}
	\frac{d^2}{d r_*^2} H(r)+\widetilde{V} H(r)=0,
\end{equation}
with the potential
\begin{equation}
	\begin{aligned}
		\widetilde{V}= & {\left[\frac{\left(r^2+a^2\right) \omega-a m}{r^2+a^2}-i G\right]^2-\frac{d G}{\mathrm{~d} r_*} } \\
		& -\frac{2 Q}{\left(r^2+a^2\right)^2}\left[r \ma^2(r-M)(1+s)(1+2 s)\right. \\
		&-K-2 i \omega r s-\frac{i r (\left(r^2+a^2\right) \omega-a m)}{\left(r^2+a^2\right)}],
	\end{aligned}
\end{equation}
where
\begin{equation}
	G=\frac{s\left[(r-M)\left(1-r^2 \ma^2\right)-r \ma^2 \Delta\right]}{\left(r^2+a^2\right)}+\frac{r Q}{\left(r^2+a^2\right)^2}.
\end{equation} Moreover, the angular equation is given by
\begin{equation}\label{angulareq}
	\frac{1}{\sin \theta} \frac{d}{d \theta}\left(\sin \theta \frac{d Y(\theta)}{d\theta}\right)+V_{\rm {(ang) }}^R(\theta) Y(\theta)=0,
\end{equation}
with
\begin{eqnarray}
V_{\rm (ang)}(\theta)&=&\frac{1-2K+s(2-\sigma_0)}{P}\nonumber\\
&&+\frac{1}{P^2}\Bigg\{ -\frac{(w\cos\theta-\sigma_0s)^2}{\sin^2\theta}\nonumber\\
&&-(z+w-4s \ma M)^2+(z\cos\theta-s\sigma_0 )^2 \nonumber\\
&&-(\ma M\cos\theta-1)^2+1-\sigma_0+\ma^2M^2\nonumber\\
&&+4s(\sigma_0-1)\cos\theta(2s\ma M-w)\Bigg\},
\end{eqnarray}
where $Y(\theta)=\sqrt{P}S(\theta)$, $z=a\omega+s M\ma$  and $w=-m+2sM\ma$.
 
Taking the Minkowski limit, $M=a=0$, the angular equation reduces to
\begin{eqnarray}
	&&\frac{1}{\sin \theta} \frac{d}{d \theta}\left(\sin \theta \frac{d Y(\theta)}{d\theta}\right)\nonumber\\
	&&\qquad+\left[s-2 K-\frac{(s \cos  \theta+m)^2}{\sin ^2 \theta}\right] Y(\theta)=0.
\end{eqnarray}
Compare this equation with the one of the spin-weighted spherical harmonics ${}_sY_{\ell m}$ \cite{Newman:1966ub}
\begin{equation}
	\begin{aligned}
		& \frac{1}{\sin \theta} \frac{\partial}{\partial \theta} \left(\sin \theta \frac{\partial}{\partial \theta} {}_sY_{\ell m}\right)\\  &\qquad+\left[\ell(\ell+1)-s^2
		 -\frac{(m+s \cos \theta)^2}{\sin ^2 \theta}\right]{ }_s Y_{\ell m}=0,
	\end{aligned}
\end{equation}
we can immediately find that
\begin{equation}\label{sep2}
	2 K=s+s^2-\ell(\ell+1).
\end{equation}
Thus the separation constant of the angular equation in the Minkowski limit is known and the QN frequencies can be obtained solely by considering the radial equation.
The radial equation \eqref{rdeq} in the Minkowski limit simplifies to
\begin{equation}
	\begin{aligned}
		& \left(1-\ma^2 r^2\right)^2 \frac{d^2 H(r)}{d r^2}-\left(1-\ma^2 r^2\right) 2 r \ma^2 \frac{d H(r)}{d r} \\
		& +\Bigg(\omega^2+\frac{2 i s \omega}{r}-\frac{\ell(\ell+1)}{r^2}+\ell(\ell+1) \ma^2  -s^2 \ma^2\Bigg) H(r)=0,
	\end{aligned}
\end{equation}
which in terms of the dimensionless quantities \eqref{scalarxeq} can be written in the form of Eq. \eqref{eqfx}
\begin{equation}
	\begin{aligned}\label{fuchss}
		\frac{d^2 H}{d x^2} & +\frac{2 x}{x^2-1} \frac{d H}{d x}+\left(-\frac{\ell(\ell+1)}{\left(x^2-1\right)^2 x^2}\right. \\
		& \left.+\frac{2 i s \widetilde{\omega}}{x\left(x^2-1\right)^2}+\frac{\widetilde{\omega}^2-s^2+\ell(\ell+1)}{\left(x^2-1\right)^2}\right) H=0.
	\end{aligned}
\end{equation}
\subsection{Fuchsian equation}
In the following we will show that Eq. \eqref{fuchss} belongs to the Fuchsian equation with three regular singular points $0$, $1$ and $-1$, which then can be easily turned into the standard form of the HDE. From \eqref{eqfx} we know
	\begin{eqnarray}
		p(x)=\frac{2 x}{x^2-1},
	\end{eqnarray}
	\begin{eqnarray}
	q(x)&=&-\frac{\ell(\ell+1)}{\left(x^2-1\right)^2 x^2}+\frac{2 i s \widetilde{\omega}}{x\left(x^2-1\right)^2}\nonumber\\
	&&+\frac{\widetilde{\omega}^2-s^2+\ell(\ell+1)}{\left(x^2-1\right)^2}.
	\end{eqnarray}
We can verify that the three points $0$, $1$ and $-1$ are the regular singular points of the equation and the infinity is an ordinary point. Furthermore, we can expand $p(x)$ and $q(x)$ around each regular singular point as \eqref{eq:px} and \eqref{eq:qx}, with the expansion coefficients given by
\begin{eqnarray}
		A_1&=&0, \nonumber\\ 
		B_1&=&-\ell(\ell+1), \\ 
		C_1&=&2 i s \widetilde{\omega},\nonumber
\end{eqnarray}
\begin{eqnarray}
		A_2&=&1, \nonumber \\
		B_2&=&-\frac{1}{4}(s-i \widetilde{\omega})^2, \\
		C_2&=&\frac{1}{4}\left(s^2-\widetilde{\omega}^2-4 i s \widetilde{\omega}+2 \ell(\ell+1)\right),\nonumber
\end{eqnarray}
and 
\begin{eqnarray}
		A_{3}&=&1, \nonumber\\ 
		B_{3}&=&-\frac{1}{4}(s+i \widetilde{\omega})^2,  \\
		C_{3}&=&\frac{1}{4}\left(-s^2+\widetilde{\omega}^2-4 i s \widetilde{\omega}-2 \ell(\ell+1)\right).\nonumber
\end{eqnarray}
Here the three regular singular points are labeled as $a_1=0$, $a_2=1$ and $a_3=-1$.
Next, from the indicial equation \eqref{eq:index} we can determine the characteristic exponents of each regular singular point, which are given by
\begin{eqnarray}
	&&\alpha_1=\ell+1,\quad \alpha_2=-\ell, \nonumber\\ &&\beta_1=\frac{1}{2}(-s+i \tilde{\omega}),\quad \beta_2=\frac{1}{2}\left(s-i \tilde{\omega}\right),\\
	&&\gamma_1=\frac{1}{2}(s+i \widetilde{\omega}),\quad
	\gamma_2=\frac{1}{2}(-s-i \widetilde{\omega}).\nonumber
\end{eqnarray}
Through the transformation \eqref{apa1}, that is
\begin{equation}
	\begin{aligned}
		& y=\frac{2x}{x+1},\\
		&H=\left(\frac{x}{x+1}\right)^{\ell+1}\left(\frac{x-1}{x+1}\right)^{\frac{1}{2}(-s+i\widetilde{\omega})} g(y),
	\end{aligned}
\end{equation}
the radial equation \eqref{fuchss} can be turned into the standard form of HDE \eqref{hyper}, with the parameters
given by \eqref{indexrel} 
\begin{equation}\label{alphabetagammaH}
	\begin{aligned}
		& \alpha=\ell+1+i\widetilde{\omega}, \\
		& \beta=\ell+1-s, \\
		& \gamma=2\ell+2.
	\end{aligned}
\end{equation}
The subsequent analysis is parallel to the scalar field case, so we shall not present all the details but the main steps here.  The requirement of being regular at the origin of the accelerating spacetime  singles out the solution of the HDE to be the hypergeometric function. To explore the behavior of the solution at the acceleration horizon, viz. $y=1$, we need to change the variable of hypergeometric function from $y$ to $1-y$, which depending on the values of the parameter $\gamma-\alpha-\beta$ can have different form, either like \eqref{noninterger} or \eqref{integer}. To satisfy the boundary condition that the field is purely outgoing near the acceleration horizon, we must impose \eqref{bound} as well, which leads to
\begin{equation}
	i \widetilde{\omega}=n+\ell+1.
\end{equation}
We obtain the same result of the QN frequency as the one in the previous section. It is interesting to note that the  spectra  are purely imaginary and independent of the spin of the perturbations. Moreover, the result of $s=-2$ matches that of the acceleration modes of gravitational perturbations of spinning C-metric \cite{Chen:2024rov} after taking the Minkowski limit.
\subsection{Direct conversion}
Similar to the analysis of the scalar field in the previous section, the radial equation \eqref{fuchss} can be converted into the standard form of the HDE through proper transformation, but the process is more complicated than the case of the scalar field. A direct thought is to make the change of variable $y=x^2$, which leads to
\begin{equation}
	\begin{aligned}
		& 4 y(y-1) \frac{d^2 H}{d y^2}+(6 y-2) \frac{d H}{d y}  +\left(-\frac{\ell(\ell+1)}{y}\right.\\
		&\quad \left.+\frac{2 i s \widetilde{\omega}}{\sqrt{y} \cdot(y-1)}+\frac{\widetilde{\omega}^2-s^2+\ell(\ell+1)}{y-1}\right) H=0.
	\end{aligned}
\end{equation}
In the case of the scalar fields $s=0$, via the change of variable $y=x^2$, the regular singular points of the equation change from  $(0, 1, -1)$ to $(0,1, \infty)$. The equation can then be turned into the form of the HDE. However, for $s\ne0$, due to the appearance of the $\sqrt{y}$ term in the equation, the point $y=0$ is not a regular singular point anymore, which makes situation even worse. Nevertheless,  the similar issue was encountered in dS spacetime  and a coordinate transformation was proposed  \cite{Suzuki:1995nh}, that is
\begin{equation}
	z=\frac{1-x}{1+x},
\end{equation}
which may be useful here. Make such  change of variable, the radial equation \eqref{fuchss} becomes
\begin{equation}
	\begin{aligned}
		& \frac{d^2 H(z)}{d z^2}+\frac{1}{z} \frac{d H(z)}{d z} \\
		& -\left(\frac{\left(s^2-\widetilde{\omega}^2\right)(z-1)^2+2 i s \widetilde{\omega}\left(z^2-1\right)+4 z \ell(\ell+1)}{4 z^2(z-1)^2}\right) \\
		&\times H(z)=0.
	\end{aligned}
\end{equation}
Now we can see that the regular singular points for this equations are $(0,1,-1)$, thus it is natural to turn this equation into the form of the HDE. Note that with the new coordinate 
$z$, the positions of the origin of the accelerating spacetime and the acceleration horizon are switched, which is not convenient for the subsequent analysis. To avoid this issue, we introduce 
\begin{equation}
	y=1-z,
\end{equation}
such that the origin is still at $y=0$ and the acceleration is at $y=1$. Then above equation becomes
\begin{equation}
	\begin{aligned}
		& \frac{d^2 H}{d y^2}+\frac{1}{y-1} \frac{d H}{d y} \\
		& -\left(\frac{\left(s^2-\widetilde{\omega}^2\right)y^2+2 i s \widetilde{\omega}y(y-2)+4 (1-y) \ell(\ell+1)}{4 y^2(y-1)^2}\right) \\
		&\times H=0.
	\end{aligned}
\end{equation}
Consider the ansatz for $H(y)$,
\begin{equation}
	H(y)=y^A(1-y)^B \widetilde{H}(y),
\end{equation}
where
\begin{equation}
	A=\left\{\begin{array}{l}
	\ell+1	, \\ \\
	-\ell	,
	\end{array} \quad B=\left\{\begin{array}{l}
		\frac{1}{2}(-s+i \widetilde{\omega}), \\ \\
	\frac{1}{2}(s-i \widetilde{\omega})	,
	\end{array}\right.\right.
\end{equation}
we get that the function $\widetilde{H}(y)$ must be solutions of the HDE
\begin{equation}
y(1-y)\frac{d^2 \widetilde{H}}{d y^2}+\left[\gamma-(\alpha+\beta+1) y\right] \frac{d \tilde{H}}{d y} -\alpha \beta \widetilde{H}=0,
\end{equation}
with the parameters given by
\begin{eqnarray}
	\alpha&=&A+B+\frac{s+i \widetilde{\omega}}{2},\nonumber \\ 
	\beta&=&A+B-\frac{s+i \widetilde{\omega}}{2}, \\
	\gamma&=&2 A.\nonumber
\end{eqnarray}
There are four equivalent choices for $A$ and $B$, the one with $A=\ell+1$ and $B=\frac{1}{2}(-s+i \widetilde{\omega})$ yields exactly the same parameters of the HDE as \eqref{alphabetagammaH}. Since the two HDEs are the same, it is expected that the QNM spectra are identical as well.  
\section{De Sitter Modes}\label{sec:level4}
In this section we will use the transformation of HDEs and Fuchsian equations with three regular singular points to calculate the dS modes to show the validity and convenience of this method.  Here we consider scalar, electromagnetic and gravitational perturbations in $D$ dimensional dS spacetime. As we will show below, our results are identical with those presented in \cite{Lopez-Ortega:2006aal}.

\subsection{Scalar field}
For simplicity in the following we only consider a massless scalar field minimally coupled to gravity living in the pure dS spacetime. The QN frequencies are embodied in the radial part of the master equation \cite{Abdalla:2002hg}, which is given by
\begin{equation}
	\begin{aligned}
		 & x(1-x) \frac{d^2 R}{d x^2}-\frac{1}{2}[(D+1) x-(D-1)] \frac{d R}{d x} \\
		& +\frac{1}{4}\left(\frac{\widetilde{\omega}^2}{1-x}-\frac{\ell(\ell+D-3)}{x}\right)R=0.
	\end{aligned}
\end{equation}
Here and in the subsequent subsections $x=r^2/L$, $ \widetilde{\omega}=\omega L$ with $L$ being the radius of the dS space and $D$ is the spacetime dimension.  We can observe that above equation have three regular singular points, which are labeled by $a_1=0$, $a_2=1$ and $\infty$. Writing the radian equation into the form of \eqref{eqfx} and then we can expand $p(x)$ and $q(x)$ around $a_1$ and $a_2$ as \eqref{eq:px} and \eqref{eq:qx}, with the expansion coefficients given by
\begin{eqnarray}
		A_1&=&\frac{D-1}{2},\nonumber \\ 
		B_1&=&-\frac{\ell(D+\ell-3)}{4}, \\ 
		C_1&=&\frac{-(D-3) \ell
			+\widetilde{\omega} ^2-\ell ^2}{4},\nonumber
\end{eqnarray}
and
\begin{eqnarray}
		A_2&=&1,\nonumber \\ 
		B_2&=&\frac{\widetilde{\omega}^2}{4}, \\ 
		C_2&=&\frac{(D-3) \ell
			-\widetilde{\omega} ^2+\ell ^2}{4}.\nonumber
\end{eqnarray}
Then from the indicial equations \eqref{eq:index} and \eqref{eq:indexinfi}, the characteristic exponents of each regular singular point, which are given by
\begin{eqnarray}
	&&\alpha_1=\frac{\ell}{2},\quad \alpha_2=\frac{1}{2}(3-D-\ell), \nonumber\\ &&\beta_1=\frac{i \widetilde{\omega}}{2},\quad \beta_2=-\frac{i \widetilde{\omega}}{2},\\
	&&\gamma_1=0,\quad
	\gamma_2= \frac{D-1}{2}.\nonumber
\end{eqnarray}
Moreover, through the transformation \eqref{apa1} the radian equation can be turned into the standard HDE with the parameters given by \eqref{indexrel}. The comparison with the analysis of the scalar field in dS spacetime \cite{Lopez-Ortega:2006aal} shows that both the transformations connecting the radial equation to the HDE and the resulting HDEs are exactly the same (see Eqs.(46)-(48) of \cite{Lopez-Ortega:2006aal}).  This clearly implies that the QNM spectra in these two cases are identical.
\subsection{Electromagnetic field}
For electromagnetic fields moving in the $D$ dimensional dS spacetime, the separation of the  equations of motion has been studied in \cite{Crispino:2000jx}. Depending on the choices of gauge of the  electromagnetic field,  there are two sets of physical solutions to the field equations, which are called physical modes I and II. Correspondingly, there are two sets of radial equations and QNM spectra \cite{Lopez-Ortega:2006aal}. 

The radial equation of physical mode I is given by
 \begin{equation}
 	\begin{aligned}
 		&  \frac{d^2 R^{(I)}}{d x^2}+\left[\frac{D+1}{2x}-\frac{1}{1-x}\right] \frac{d R^{(I)}}{d x} \\
 		& +\frac{1}{4}\left[\frac{\widetilde{\omega}^2}{x(1-x)^2}-\frac{(\ell-1)(\ell+D-2)}{x^2(1-x)}-\frac{3(D-2)}{x(1-x)}\right] \\
 		&\quad \times R^{(I)}=0.
 	\end{aligned}
 \end{equation}
We can observe that this equation has three regular singular points $a_1=0$, $a_2=1$ and $\infty$. As before, we can write this equation into the form of \eqref{eqfx} and then  expand $p(x)$ and $q(x)$ around $a_1$ and $a_2$ as \eqref{eq:px} and \eqref{eq:qx}, with the expansion coefficients given by
\begin{eqnarray}
	A_1&=&\frac{D+1}{2},\nonumber \\ 
	B_1&=&-\frac{1}{4} (\ell -1) (D+\ell -2), \\ 
	C_1&=&\frac{1}{4} \left(-D (\ell +2)+\widetilde{\omega}^2-\ell ^2+3 \ell +4\right),\nonumber
\end{eqnarray}
and
\begin{eqnarray}
	A_2&=&1,\nonumber \\ 
	B_2&=&\frac{\widetilde{\omega}^2}{4}, \\ 
	C_2&=&\frac{1}{4} \left(D (\ell +2)-\widetilde{\omega}^2+\ell ^2-3 \ell -4\right).\nonumber
\end{eqnarray}
Then from the indicial equations \eqref{eq:index} and \eqref{eq:indexinfi}, the characteristic exponents of each regular singular point are given by
\begin{eqnarray}
	&&\alpha_1=\frac{\ell -1}{2},\quad \alpha_2=\frac{1}{2} (-D-\ell +2), \nonumber\\ &&\beta_1=\frac{i \widetilde{\omega}}{2},\quad \beta_2=-\frac{i \widetilde{\omega}}{2},\\
	&&\gamma_1=\frac{D-2}{2},\quad
	\gamma_2=\frac{D-2}{2}.\nonumber
\end{eqnarray}
We can compare above results with those in \cite{Lopez-Ortega:2006aal} and find that both the transformations connecting the radial equation to the HDE and the resulting HDEs are exactly the same (see Eqs.(7)-(10) of \cite{Lopez-Ortega:2006aal}). This means the resulting QNM spectra would be identical.

The radial equation of physical mode II is given by
\begin{equation}
	\begin{aligned}
		&  \frac{d^2 R^{(II)}}{d x^2}+\frac{D (1-x)+x-3}{2 (1-x) x}\frac{d R^{(II)}}{d x} \\
		& +\frac{1}{4 x (1-x)}\left[\frac{\widetilde{\omega} ^2}{1-x}-\frac{(\ell +1) (D+\ell -4)}{x}\right] R^{(II)}=0.
	\end{aligned}
\end{equation}
Similar to the physical mode I, this equation have three regular singular points $a_1=0$, $a_2=1$ and $\infty$. As before, we can write this equation into the form of \eqref{eqfx} and then  expand $p(x)$ and $q(x)$ around $a_1$ and $a_2$ as \eqref{eq:px} and \eqref{eq:qx}, with the expansion coefficients given by
\begin{eqnarray}
	A_1&=&\frac{D-3}{2},\nonumber \\ 
	B_1&=&-\frac{1}{4} (\ell +1) (D+\ell -4), \\ 
	C_1&=&\frac{1}{4} \left(-D (\ell +1)+\widetilde{\omega}^2-\ell ^2+3 \ell +4\right),\nonumber
\end{eqnarray}
and
\begin{eqnarray}
	A_2&=&1,\nonumber \\ 
	B_2&=&\frac{\widetilde{\omega}^2}{4}, \\ 
	C_2&=&\frac{1}{4} \left(D (\ell +1)-\widetilde{\omega}^2+\ell ^2-3 \ell -4\right).\nonumber
\end{eqnarray}
Then from the indicial equations \eqref{eq:index} and \eqref{eq:indexinfi}, the characteristic exponents of each regular singular point are given by
\begin{eqnarray}
	&&\alpha_1=\frac{\ell +1}{2},\quad \alpha_2=\frac{1}{2} (-D-\ell +4), \nonumber\\ &&\beta_1=\frac{i \widetilde{\omega}}{2},\quad \beta_2=-\frac{i \widetilde{\omega}}{2},\\
	&&\gamma_1=0,\quad
	\gamma_2=\frac{D-3}{2}.\nonumber
\end{eqnarray}
Similar to the physical mode I, both the transformations connecting the radial equation to the HDE and the resulting HDEs are exactly the same as those in \cite{Lopez-Ortega:2006aal} (see Eqs.(22)-(24) therein). This means the resulting QNM spectra would be same.

\subsection{Gravitational perturbations}
In dimensions higher than four,  the gravitational perturbations of $D$ dimensional dS spacetime are grouped in three types, according to their tensorial behavior on the $(D-2)$ sphere: tensor, vector and scalar types \cite{Kodama:2003kk}. In contrast, there is only tensor type for gravitational perturbations in four dimension. The radial equations of these types of perturbations in this background can be uniformly expressed as \cite{Natario:2004jd}
\begin{equation} 
	\begin{aligned}
		&  \frac{d^2 R_{G}}{d x^2}+\frac{1-3 x}{2 x(1-x)}\frac{d R_{G}}{d x} \\
		& +\frac{1}{4 x (1-x)^2}\left[\widetilde{\omega}^2+\widetilde{\alpha}  (1-x)-\frac{\widetilde{\beta} (\widetilde{\beta} +1) (1-x)}{x}\right] \\
		&\quad \times R_G=0,
	\end{aligned}
\end{equation}
where $\widetilde{\alpha}$ and $\widetilde{\beta}$ are introduced by
\begin{equation}
	\widetilde{\alpha}=	\left\{\begin{array}{l}
	\frac{(D-2) D}{4} ,\quad\text{tensor type},\\ 
		\\
		\frac{(D-4)(D-2)}{4},\quad\text{vector type}, \\ \\
		
		\frac{(D-6)(D-4)}{6},\quad\text{scalar type},
	\end{array}\right.
\end{equation}
and
\begin{equation}
	\widetilde{\beta}=\frac{2\ell+D-4}{2}.
\end{equation}
This equation has three regular singular points $a_1=0$, $a_2=1$ and $\infty$. As before, we can write this equation into the form of \eqref{eqfx} and then  expand $p(x)$ and $q(x)$ around $a_1$ and $a_2$ as \eqref{eq:px} and \eqref{eq:qx}, with the expansion coefficients given by
\begin{eqnarray}
	A_1&=&\frac{1}{2},\nonumber \\ 
	B_1&=&-\frac{1}{4} \widetilde{\beta}  (\widetilde{\beta} +1), \\ 
	C_1&=&\frac{1}{4} \left(\widetilde{\alpha} -\widetilde{\beta} ^2-\widetilde{\beta} +\omega ^2\right),\nonumber
\end{eqnarray}
and
\begin{eqnarray}
	A_2&=&1,\nonumber \\ 
	B_2&=&\frac{\widetilde{\omega}^2}{4}, \\ 
	C_2&=&-\frac{1}{4} \left(\widetilde{\alpha} -\widetilde{\beta} ^2-\widetilde{\beta} +\omega ^2\right).\nonumber
\end{eqnarray}
Then from the indicial equations \eqref{eq:index} and \eqref{eq:indexinfi}, the characteristic exponents of the three regular singular points are given by
\begin{eqnarray}
	&&\alpha_1=\frac{1+\widetilde{\beta}}{2},\quad \alpha_2=-\frac{\widetilde{\beta}}{2}, \nonumber\\ &&\beta_1=\frac{i \widetilde{\omega}}{2},\quad \beta_2=-\frac{i \widetilde{\omega}}{2},\\
	&&\gamma_1=\frac{1}{4} \left(1+\sqrt{4 \widetilde{\alpha}+1}\right),\quad
	\gamma_2=\frac{1}{4} \left(1-\sqrt{4 \widetilde{\alpha}+1}\right).\nonumber
\end{eqnarray}
Again, both the transformations connecting the radial equation to the HDE and the resulting HDEs are exactly the same as those in \cite{Lopez-Ortega:2006aal} (see Eqs.(34)-(36) therein). Therefore, the QNM spectra in these two works would be identical.
\section{Summary}\label{sec:level5}
In this work, we have investigated the QNMs of perturbations with spin $s\leq2$ in pure accelerating spacetime to explore the origin of the acceleration modes recently discovered for scalar and gravitational perturbations of accelerating BHs \cite{Destounis:2020pjk,Xiong:2023usm,Chen:2024rov}. The perturbation equations were obtained by taking the Minkowski limit of the master equations describing various types of perturbations of accelerating BHs. We have employed two methods to get the QNM spectra by solving the radial part of the perturbation equations. One is via change of variable to convert the radial equation into the form of the standard HDEs. This works well for scalar perturbations as the case for dS mode in pure dS spacetime \cite{Lopez-Ortega:2006aal}, but for other types of perturbations the operation is more complicated. The alternative method involves using the simple connection between Fuchsian equations with three regular singular points--to which the radial perturbation equations belongs to-- and HDEs. These two methods were shown to produce the same QNM spectra.

It is interesting to find that the resulting QNM spectra of the pure accelerating spacetime are purely imaginary and independent of the spin of the perturbations, and match those of the acceleration modes of accelerating BHs after taking the Minkowski limit. This implies that the acceleration modes actually originate from the pure accelerating spacetime and the appearance of BHs would deform the spectra. The situation is very similar to the one of the dS modes \cite{Jansen:2017oag, Cardoso:2017soq,Destounis:2018qnb, Konoplya:2022xid}, which were first calculated in pure dS spacetime and then were identified in the scalar QNMs of Schwarzschild-dS BHs \cite{Jansen:2017oag}. In addition, we have also applied the second method to the calculation of QNMs of scalar, electromagnetic and gravitational perturbations in $D$ dimensional dS spacetime and the results are identical to those in \cite{Lopez-Ortega:2006aal}, which verifies the validity of this method.

Given that acceleration modes are closely linked to the acceleration parameter and exhibit robustness to BH parameters, it would be intriguing to explore the observability of these modes. This could shed light on the acceleration of BHs within GW signals, potentially aiding in the detection of moving and accelerating BHs.

\begin{acknowledgments}
We would like to thank Wei Xiong for his kind help at the initial stage of this work. The work is in part supported by NSFC Grant
No.12205104, ``the Fundamental Research Funds for the Central Universities'' with Grant No.  2023ZYGXZR079, the Guangzhou Science and Technology Project with Grant No. 2023A04J0651 and the startup funding of South China University of
Technology. 
\end{acknowledgments}

\appendix
\section{Some basics of Fuchsian equations}\label{appdixFeq}
 Given a second-order linear ordinary differential equation
\begin{equation}\label{eqfx}
	f^{\prime \prime}(x)+p(x) f^{\prime}(x)+q(x) f(x)=0,
\end{equation}
where $a_r$ $(r=1,2, \cdots, n)$ and $\infty$ are regular singular points of this equation. Besides, since $a_r$ is a regular singular point, it is also the first-order pole of $p(x)$ at most, we have
\begin{equation}\label{eq:px}
	p(x)=\sum_{r=1}^n \frac{A_r}{x-a_r}+\varphi(x)
\end{equation}
where $A_r$ is the residue of  $p(x)$ at $x=a_r$, $\varphi(x)$ is analytical in complex plane. When $x\rightarrow\infty$,\ $\varphi(x)$ approaches to 0 since\ $x=\infty$ is the regular singular point of the equation, which implies $\varphi(x)$ can be set to zero.

Based on similar analysis, $q(x)$ can be expanded as
\begin{equation}\label{eq:qx}
	q(x)=\sum_{r=1}^n\left\{\frac{B_r}{\left(x-a_r\right)^2}+\frac{C_r}{x-a_r}\right\},
\end{equation}
with
\begin{equation}
	\sum_{r=1}^n C_r=0.
\end{equation}
When considering series solution of Eq. \eqref{eqfx}, the coefficient of the lowest power of $x$ satisfies the so-called {\em indicial equation}. The roots of the indicial equation determine the {\em characteristic exponents}, which in turn dictate the form of the solution. For each regular singular point $a_r$ ($a_r\ne\infty$) of Eq. \eqref{eqfx}, the corresponding two characteristic exponents satisfy the following indicial equation
\begin{equation}\label{eq:index}
	\rho^2+\left(A_r-1\right) \rho+B_r=0 \quad(r=1,2, \cdots, n).
\end{equation}
Moreover, if the point at infinity $x=\infty$ is a regular singular point, its characteristic exponents obey the equation
\begin{equation}\label{eq:indexinfi}
	\rho^2+\left(1-\sum_{r=1}^n A_r\right) \rho+\sum_{r=1}^n\left(B_r+a_r C_r\right)=0.
\end{equation}
From these two equations, we can obtain an important constraint in this case 
\begin{equation}\label{eq:sumindex}
	\sum {\rm roots\;of\;all\;indicial\;equations}=n-1.
\end{equation}
For example, for the HDE, the sum of all characteristic exponents is 1, since $n=2$.

On the other hand, if $x =\infty$ is an ordinary point rather than a singular point of the Eq. \eqref{eqfx}, the parameters $A_r$, $B_r$ and $C_r$ satisfy the following equations, 
\begin{eqnarray}\label{xinftyeq}
	\sum_{r=1}^n A_r&=&2,\nonumber \\
	\sum_{r=1}^n C_r&=&0,\nonumber\\
	\sum_{r=1}^n\left(B_r+a_r C_r\right)&=&0,\nonumber\\
	\sum_{r=1}^n\left(2 a_r B_r+a_r^2 C_r\right)&=&0.
\end{eqnarray}
For Fuchsian equations with three regular singular points, say $a$, $b$ and $c$ ($\neq\infty$, so $\infty$ is an ordinary point), from above formulas we can write the equations in the following form
\begin{equation}
	\begin{aligned}\label{fuchs3}
		&\frac{d^2 f}{d x^2}  +\left\{\frac{1-\alpha_1-\alpha_2}{x-a}+\frac{1-\beta_1-\beta_2}{x-b}+\frac{1-\gamma_1-\gamma_2}{x-c}\right\} \frac{d f}{d x} \\
		& +\left\{\frac{\alpha_1 \alpha_2(a-b)(a-c)}{x-a}+\frac{\beta_1 \beta_2(b-c)(b-a)}{x-b}\right. \\
		& \left.+\frac{\gamma_1 \gamma_2(c-a)(c-b)}{x-c}\right\} \times \frac{f}{(x-a)(x-b)(x-c)}=0,
	\end{aligned}
\end{equation}
where ($\alpha_1$, $\alpha_2$), ($\beta_1$, $\beta_2$) and ($\gamma_1$, $\gamma_2$) are the corresponding characteristic exponents of the regular singular points $a$, $b$ and $c$. From \eqref{eq:index} and \eqref{xinftyeq} we immediately find
\begin{equation}\label{alphasbeta1}
	\alpha_1+\alpha_2+\beta_1+\beta_2+\gamma_1+\gamma_2=1.
\end{equation}
Through the following transformation 
\begin{equation}\label{apa1}
	\begin{aligned}
		y & =\frac{(b-c)(x-a)}{(b-a)(x-c)}, \\
		f& =\left(\frac{x-a}{x-c}\right)^{\alpha_1}\left(\frac{x-b}{x-c}\right)^{\beta_1} g,
	\end{aligned}
\end{equation}
Eq. \eqref{fuchs3} can be turned into the standard form of HDE, 
\begin{equation}\label{hyper}
	y(1-y)\frac{d^2 g}{d y^2}+\left[\gamma-(\alpha+\beta+1) y\right] \frac{d g}{d y} -\alpha \beta g=0,
\end{equation}
whose parameters are related to those of the Fuchsian equations by
\begin{equation}
	\begin{aligned}\label{indexrel}
		& \alpha=\gamma_1+\alpha_1+\beta_1, \\
		& \beta=\gamma_2+\alpha_1+\beta_1, \\
		& \gamma=1+\alpha_1-\alpha_2.
	\end{aligned}
\end{equation}
The detailed derivation of above formulas using Riemann's $P$-functions can be found in \cite{wang1989special}.


\bibliographystyle{apsrev4-1}
%

\end{document}